\documentclass[draft]{article} 
\usepackage{latexsym,amssymb,amsmath,amscd,amsthm,amsxtra}
\newtheorem{Th}{Theorem}
\newtheorem{lem}{Lemma}
\newtheorem{co}{Corollary}

\newcommand{\rr}{\mathbb {R}^n}
\newcommand{\N}{\mathbb {N}}

\title{\Large{\bf Statistical Thermodynamics of General Minimal Diffusion\\
Processes: Constuction, Invariant Density,\\
Reversibility and Entropy Production\footnote{A condensed version of
this paper will be published by the {\it Journal of Statistical 
Physics.}}}}

\author{HONG QIAN\thanks{School of Mathematical Sciences, Peking 
University, Beijing, 100871, P.R.C.; Department of Applied Mathematics, 
University of Washington, Seattle, Washington, 98195-2420
(qian@amath.washington.edu).},\ \
MIN QIAN\thanks{School of Mathematical Sciences, 
Peking University, Beijing, 100871, P.R.C.}\ \ 
and XIANG TANG\thanks{School of Mathematical Sciences, 
Peking University, Beijing, 100871, P.R.C.; Department of
Mathematics, University of California, Berkeley, CA 94720-3840
(xtang@math.berkeley.edu).}}

\begin{document}
\maketitle
\large 
\begin{abstract}
The solution to nonlinear Fokker-Planck equation is constructed
in terms of the minimal Markov semigroup generated by the equation.
The semigroup is obtained by a purely functional analytical method
via Hille-Yosida theorem.  The existence of the positive invariant 
measure with density is established and a weak form of Foguel 
alternative proven.  We show the equivalence among self-adjoint
of the elliptic operator, time-reversibility, and zero entropy 
production rate of the stationary diffusion process.  A 
thermodynamic theory for diffusion processes emerges. 

\end{abstract}

\section*{\bf 0\ \ \ Introduction}
\noindent

	Diffusion processes, as an important part of 
statistical mechanics, are models for many equilibrium 
and nonequilibrium phenomena.   It is widely considered
as a phenomenological approach to systems with
fluctuations; however, its relationship to nonequilibrium 
thermodynamics is not clear.  In recent years, motivated by
work on biological macromolecules which convert chemical 
energy into mechanical work (molecular motors) \cite{Q1,FK}, 
it becomes evident that a thermodynamic formalism, 
both for equilibium and more importantly nonequilibrium, 
can be developed from a diffusion theory of macromolecules 
in an ambient fluid at constant temperature \cite{Q3,Q6}.  
This is a natural extension of the dynamic theory 
of synthetic  polymers which are passive molecules \cite{DE}. 
Molecular motors are nano-scale devices, driven and
operating under nonequilibrium steady-state with heat
dissipation \cite{Q2}.  

	The central elements in the new development are
the heat dissipation and the entropy production
\cite{Q3}.  The essential difference between an equilibrium  
polymer and a molecular motor is that the former has 
zero heat dissipation and entropy production while for
the latter they are positive.  Introducing these two
quantities into a diffusion process makes the 
stochastic theory a thermodynamic one with the
first and the second laws, as well as Onsager's theory \cite{O},
as logical outcomes \cite{Q6}. 

	Heat dissipation and entropy production also play 
important roles in the mathematical formulations for the 
stationary nonequilibrium steady-state (NESS) in a computer 
simulation of driven fluids \cite{ECM}.  Numerical
observations have led to a surge of mathematical analysis  
of NESS from a dynamical-system point of view \cite{R};
Lebowitz and Spohn also developed a theory from a 
stochastic-process standpoint \cite{LS}.  Later, the mathematical
relationship between the entropy production in the diffusion 
theory and that in the axiom-A system has been established
\cite{JQQ}, and an intimate relationship between Lebowitz
and Spohn's approach and the diffusion-process based thermodynamics 
has also been discussed \cite{Q6}. 

	As the foundation for the new statistical thermodynamic 
theory, the mathematical task is to firmly establish the relation 
between time reversibility and vanishing of the entropy production
for general diffusion processes.  A technical difficulty to be 
overcome is to rigorously construct a stationary diffusion process
from a given stochastic models in the form of a stochastic 
differential equation 
\begin{equation}
         \frac{{\rm d}x}{{\rm d}t} = b(x) + \Gamma \xi(t), 
				\hspace{1.5cm} x \in \rr
\label{SDE} 
\end{equation}
or its corresponding Fokker-Planck equation
\begin{eqnarray}
    \frac{\partial u}{\partial t} = {\cal L}^{*} u(t,x)
	   &\triangleq& \nabla\cdot\left(\frac{1}{2} A(x) \nabla u 
			+ b(x) u\right),
       		\hspace{.5cm} (A = \Gamma\Gamma^{T}),
\label{FPE}\\
                    u(0,x) &=& f(x),
\label{cauchy}
\end{eqnarray}
where ${\cal L}^*$ denotes the adjoint of operator ${\cal L}$.
Constructing the diffusion process is usually 
accomplished by a probabilistic method based on maringales
\cite{SV}.  Here we provide an alternative purely functional 
analytical approach, which enables us to rigorously establish 
the self-adjoint (symmetric and maximum on an appropriate 
Hilbert space) generator for reversibility.  This approach 
is also more accessible for readers familiar with the 
mathematical physics of quantum mechanics. 

	In this paper we study the diffusion process defined 
by the nonlinear stochastic differential equation (\ref{SDE}) in 
which $\Gamma$ is a nonsingular matrix and $\xi (t)$ is the 
``derivative'' of a $n$-dimensional Wiener process.  This equation 
has wide applications in science and engineering as a model for 
continuous stochastic movement.  One standard method for attacking
this equation is to find the fundamental solution to the Cauchy 
problem of the related Fokker-Planck (Kolmogorov forward) equation 
(\ref{FPE}), which defines the transition probability 
$\tilde{p}(t,x,{\rm d}y)$ on the entire $\rr$.  
Unfortunately in the theory of partial differential equations the 
existence and uniqueness of the fundamental solution to Eq. \ref{FPE} 
imposes very restrictive conditions, i.e., boundness on the coefficients 
$A(x)$ and $b(x)$ \cite{LM}. Most of the interesting applications of 
Eq. \ref{SDE} could not meet the required conditions.  One 
eminent example is the Ornstein-Uhlenbeck process associated with
Eq. \ref{SDE} with linear $b(x)$. 

	For nonlinear $b(x)$ defined on entire $\rr$, in general the 
uniqueness of the solution to Eq. \ref{FPE} does not hold true.  To 
circumvent this predicament, we shall directly construct the minimal 
semigroup generated by the Fokker-Planck equation by a purely
functional analytical method instead of the traditional probabilistic
one \cite{SV}.  The existence of a family of transition 
functions satisfying Kolmogorov-Chapman 
equation then follows.  By finding the invariant functional in the 
non-zero limit case, we obtain an invariant probability density.  
Hence by this approach we obtain a weak Foguel alternative and a 
stationary Markov process as a solution to Eq. \ref{SDE}.  This
approach is new, even though a part of the mathematics has been in 
the Chinese literature \cite{QM1,QM2}.  We present some of the 
relevant results here for the completeness for english audiences. 

 	In mathematics, \cite{QQ1} gave the first rigorous result on 
irreversibility and entropy production in the case of discrete-state 
Markov chains.  A comprehensive treatment of this case has been published 
\cite{Ka}.  For a diffusion process with bounded coefficients $A(x)$ and 
$b(x)$, related results were anounced in \cite{QQ2,QQG} where Girsanov 
formula could be used in the proof.  This approach, however, is not
valid for the case of unbounded $A(x)$ and $b(x)$, on $\rr$, which is 
addressed here.  For linear $b(x)$ in Eq. \ref{SDE}, the mathematical 
task is significantly simplified and the diffusion processes are 
also Gaussian.  The linear case is intimately related to Onsager's 
theory of irreversibility \cite{Q3}.  

	In the following, we assume:

1) $A(x)$ = $\{a_{ij}(x)\}$, $b(x)$ = $\{b_{j}(x)\}$  are smooth;

2) $\nabla\cdot b(x)\geq \mu_0$, where $\mu_0$\ is a constant;

3) Uniformly elliptic condition
\[
\sum_{i,j=1}^{n}a_{ij}(x)\xi_{i}\xi_{j} \geq r\sum_{i=1}^{n}\xi_{i}^{2}
		\hspace{1.5cm} \forall \xi \in \rr
\]
where $r$ is a positive constant.  

	The paper is organized as follows.  In Section 1, we 
motivate the mathematical definition of entropy production rate
and time-reversibility by a heuristic thermodynamic analysis,
based on the concept of entropy and the equation for entropy
balance \cite{NP}. 
In Section 2, we first construct the resolvent operators. 
Then by applying Hille-Yosida theorem, we obtain the semigroup 
generated by the solution to Kolmogorov forward (Eq \ref{FPE})
and backward equations in appropriate Banach spaces $\hat{C}(\rr)$ 
and $\widetilde{C}(\rr)$ respectively.
In Section 3, we show that the semigroup has a family of transition 
functions satisfying the Kolmogorov-Chapman equation, and prove the 
existence of the invariant probability density. 
In Section 4, The equivalence among reversibility, zero 
entropy production rate, and symmetricity of operator ${\cal L}$ 
is established for general minimal diffusion processes.

\section{The Thermodynamic Formalism of Diffusion Processes}
\noindent

	This section is heuristic. The most important concepts in 
thermodynamics are entropy and heat.  The thermodynamic formalism of 
diffusion processes provides mathematical definitions for these two 
quantities.  The entropy has the well-known definition $e[P]$ =
$-\int _{\rr} P(t,x)\log P(t,x){\rm d}x$ which is a 
functional of the probability density $P(t,x)$, the solution to
Eq. \ref{FPE}.  Let's introduce probability flux
\[ 	{\cal J}=-\frac {1}{2}A(x)\nabla P(t,x)-b(x)P(t,x).    \] 
The concept of heat is a microscopic one, hence it is an
functional of the diffusion trajectory $x(t)$: $W(t)$ = 
$2\int_0^t (A^{-1}(x)b(x(s))\circ dx(s)$ where 
$\circ$ denotes the Stratonovich integral \cite{LS}.  
Therefore, the mean heat dissipation rate (hdr) is the expectation 
$\lim_{t\rightarrow\infty} E[W(t)/t]$ = 
$\int_{\rr}2b(x)A^{-1}(x){\cal J}dx$.  
For system with detailed balance, $W(t)$ is bounded almost surely. 
Otherwise, it is not. The logarithmic generating function of $W(t)$,
\[     \lim_{t\rightarrow\infty} -\frac{1}{t}
		\log E\left[ e^{-\lambda W(t)} \right]	 \]
is convex and possesses certain symmetry, which generalizes that 
hdr being nonnegative in stationary state \cite{LS}.

 	The rate of the increase of entropy is then
$\dot{e}[P]$ = epr$-$hdr, where 
\begin{equation}
epr=\int_{\rr} 2\left({\cal J} A^{-1}(x) {\cal J}\right) 
		P^{-1}(t,x){\rm d}x, \qquad
hdr=\int_{\rr}2A^{-1}b(x)\cdot{\cal J}{\rm d}x.
\label{eprhdr}
\end{equation}

	If the force $\frac{1}{2}A(x)b(x)$  = $-\nabla U(x)$ is 
conservative, then one can also introduce free energy 
$h[P]=u[P]-e[P]$
in which $u[P]=\int_{\rr} U(x)dx$ is the internal energy and 
$\dot{u} =-hdr$.  Then $\dot{h} = -epr \ge 0$ with the equality hold ture
for the stationary process: This is the second law of thermodynamics 
applied to isothermal processes with canonical ensembles.  

	3. For nonconservative $F(x)$ without a potential, the
free energy can not be defined.  In this case, one writes $F(x)$ 
in terms of Helmholtz-Hodge decomposition: 
$F(x)=-\nabla\phi + \gamma(x)$ where the $\gamma$ is directly related 
to the circulation of the irreversible process \cite{QQ1,Q1,QW}.   

In the derivation, we used Eq. \ref{FPE} and integration by part, assuming 
no flux boundary condition and the matrix $A$ being nonsingular.  It's 
meaningful from thermodynamics point of view to identify the first term in 
Eq. \ref{eprhdr} with the entropy production rate, and second term with 
the heat dissipation rate which is the product of force 
$F(x)=(2A^{-1}b(x))$ and flux $({\cal J})$. The force in turn is the 
product of frictional coefficient $(2A^{-1})$ and velocity $b(x)$. In 
a time independent stationary state, $\dot{e}=0$, and the entropy 
production is balanced by the heat dissipation.  The following remarks 
are in order.

The entropy production rate and time-reversibility.\\
{\bf Definition 1}  The entropy production rate, epr, of a stationary 
diffusion process defined by Eq. (\ref{SDE}) is
\[
	\frac{1}{2}\int(\nabla \log P(t,x)+2A^{-1}b(t,x))^{T}A(\nabla 
		\log P(t,x)+2A^{-1}b(x))P(t,x){\rm d}x.
\]
In the stationary case, $P(t,x)= w(x)$.

\noindent
{\bf Definition 2}  
A stationary stochastic process $\{x(t);t \in {\mathbb R}\}$\ is 
time-reversible if $\forall m \in \N$ and every 
$t_1,t_2,\cdots,t_{m} \in {\mathbb R} $, the joint probability 
distribution 
\[
P(x(t_1),x(t_2),\cdots, x(t_{m}))=P(x(-t_1),x(-t_2),\cdots,x(-t_{m})).
\]

\section{The Minimal Semigroup}
\noindent

	We denote
\[ \begin{array}{ll}
   C(\mathbb{R}^n)     & = \{\textrm{bounded continuous function} f(x)\}, \\
   C_0({\mathbb{R}^n}) & = \{f\in C(\mathbb{R}^n) \mid
		         \lim_{|x|\rightarrow\infty} f(x)=0\ 
			 \textrm{uniformly} \},                  \\   
   \Vert f(x) \Vert    & = \sup_{x \in \mathbb{R}^n}|f(x)|,      \\
   \Vert\cdot\Vert     & \textrm{is the norm on } C(\mathbb{R}^n)\  	
			 \textrm{and } C_0(\mathbb{R}^n).
\end{array}
\]

	The conjugate of the Fokker-Planck equation (\ref{FPE})
is the Kolmogorov backward equation: 
\begin{equation}
   \frac{\partial u(t,x)}{\partial t} = {\cal{L}} u(t,x) =
     \frac{1}{2} \sum_{i,j=1}^n a_{ij}(x)
     \frac{\partial^2 u}{\partial x_i \partial x_j}
     - \sum_{i=1}^{n}b_i(x) \frac{\partial u}{\partial x_{i}},
          \hspace{0.5cm} (t > 0, x \in \mathbb{R}^n)
\label{KBE}
\end{equation}
For the solutions to Eqs. \ref{FPE} and \ref{KBE}, we have the 
following theorem:
\begin{Th}
If the coefficients of Eq. \ref{KBE} satisfy assumptions 1) and 3), then 
there exists a Banach space $\hat{C}(\mathbb{R}^n)$ satisfying 
$C_0(\rr)$ $\subset$ $\hat{C}(\rr)$ $\subset$ $C(\rr)$, and the semigroup 
generated by the solution to the Cauchy problem (\ref{FPE}) and 
(\ref{cauchy}) with initial data $f(x)$ exists in $\hat{C}(\rr)$.
\end{Th}

The proof of Theorem 1 are divided into four steps:

(i) $\forall n \in \N$ (the positive integers), on the bounded 
domain $B_n$ $\triangleq \{x \in \rr\mid |x| \leq n\}$, solve the 
elliptic equation; 

(ii) $\forall \lambda > 0$, construct the resolvent operator
$R(\lambda):C(\rr)\rightarrow C(\rr)$, satisfying 
$\forall f \in\ C(\rr), (\lambda-{\cal L})R(\lambda) f$ = $f$ in 
$\rr$ and $\Vert R(\lambda)\Vert \leq \frac{1}{\lambda}$;

(iii) Using $R(\lambda)$, define a Banach space $\hat{C}(\rr)$, 
satisfying $C_0(\rr)$ $\subset$ $\hat{C}({\rr})$ $\subset$ $C(\rr)$; 

(iv) The resolvent operators of ${\cal L}$ in $\hat{C}(\rr)$
satisfy the conditions of Hille-Yosida theorem.  Hence we obtain the 
semigroup generated by ${\cal L}$ which is the solution to the Cauchy 
problem (\ref{cauchy}) and (\ref{KBE}).

\subsection{\large Elliptic Equation in a Bounded Domain}

\begin{lem}
\label{lemma1}
$\forall n \in \N$ and $\forall g \in C_0(B_n)\triangleq
\{f \in C(B_n) \mid\  f|_{\partial B_n}=0\}$; the elliptic equation
\begin{eqnarray}
		(\lambda - {\cal {L}})u &=& g  
\label{eq5}\\
		u(x)|_{\partial B_n} &=& 0     
\label{eq6}
\end{eqnarray}
has  a unique solution f $ \in  C^2(B_n) \cap C_0(B_n)$.
\end{lem}

\begin{proof} 
This is a well known classic result and there is a purely functional 
analytic proof \cite{Y}.  Here we give a sketch.  $a_{ij}(x)$ and 
$b_i(x)$ are all bounded and smooth on the bounded domain $B_n$.  
By a set of inequalities given in \cite{Y}, p.420, by Riese' 
representation theorem and Lax-Milgram theorem, it was shown 
that when $\lambda \geq \mu_0$,  a sufficiently large 
constant, the Eqs. (\ref{eq5}) and (\ref{eq6}) have a solution 
$f \in \mathbb{H}^2(B_{n})$, where $\mathbb{H}^2$ is the 
Sobolev space.  Because $\partial B_n \in C^{\infty}$,\ the  weak 
solution is just the strong one according to Friedrichs-Lax-Nirenberg
theory.  Thus $ f \in C^2(B_n) $.\ So when $\lambda > \mu_0$,
the theorem is proved.

	Now for $\lambda \leq \mu _0.$ First we choose a 
$\lambda _0 > \mu _0$.\ According to the foregoing, for 
$\lambda=\lambda_0$\ there exists the solution of (\ref{eq5}) and 
(\ref{eq6}), which is denoted by $f=\tilde{R}(\lambda _0)g$.  Since 
${\cal L}$ is elliptic, $({\cal {L}}\tilde{R}(\lambda_0)g)(x_0)\leq 0$
where $x_0$ is the maximum value point of $\tilde{R}(\lambda_0)g$.  
From this we can easily prove that 
\[
\tilde{R}(\lambda _0):\ \ C_0(B_n) \rightarrow C_0(B_n) \cap C^2 (B_n),\ \ \ \Vert \tilde{R}(\lambda _0) \Vert \leq \frac {1}{\lambda _0}.
\]   
Thus, when $|\lambda_0-\lambda|\leq\frac{1}{\Vert\tilde{R}(\lambda _0)\Vert}$,
$\tilde{R}(\lambda)=\sum_{n=0}^{\infty}(\lambda_0-\lambda)^{n} (\tilde{R}(\lambda_0))^{n+1}$ is well defined.
$\lambda_0-{\cal L}$ can always be extended to being a close operator. 
Therefore, when 
$|\lambda-\lambda_0|\leq\frac{1}{\Vert\tilde{R}(\lambda _0)\Vert}$, 
\[
( \lambda - { \cal {L}}) \tilde{R}(\lambda) g = \left[( {\lambda_0} - {\cal{ L}}) - ({\lambda_0} - \lambda )\right]\left[\tilde{R}(\lambda_0)+
\cdots  \right] g=g.
\]
Hence, $\tilde{R}(\lambda)g$ is the solution of (\ref{eq5}) and (\ref{eq6}).
Finally, since  $\Vert \tilde{R}(\lambda_0) \Vert \leq \frac {1}{\lambda_0}$, we have $ |\lambda - \lambda_0| < | \lambda _0 |< \frac {1}
{\Vert \tilde{R}(\lambda_0) \Vert }$, for $ \forall 0 < \lambda < \lambda _0$.\ Thus, for  $ \forall 0<\lambda<\lambda _0,\ g \in C_0(B_n)$,\
the solution of the Eqs. (\ref{eq5}) and (\ref{eq6}) exists.

By the maximum principle of elliptic equation, we could conclude 
that the solution is unique. 
\end{proof}

\subsection{\large Resolvent Operators}

	First, choose a sequence of smooth functions 
$g_n :\rr \rightarrow \mathbb {R}$
\[
      g_n(x) \triangleq \frac {\int_{|x|^2}^{\infty} f_n(t)dt}
	{\int_{-\infty}^{\infty} f_n(t)  dt}  
\]
where
\[
    f_n (x) \triangleq \left\{
\begin{array}{ll}
     e^{\frac {1}{\left(x-n^2\right) \left(x-\left(n
	-\frac{1}{2}\right)^2\right)}} 
	& \left(n-\frac{1}{2}\right)^2 \leq |x| \leq n^2\\
	0  & \textrm{else}
\end{array}\right.
\]
We can show  that $g_n \in C_0^{\infty}(\rr)$, and 
$0\leq g_n\leq 1$, $g_n\Big|_{B_{\left(n-\frac{1}{2}\right)}}=1$, 
$g_n\Big|_{(B_{n})^{c}}=0$.
Then, \ $\forall \lambda >0$, using $ \{g_n\}_{n=1}^{\infty}$, we define 
a sequence of linear operators $R_n (\lambda )$ on $C(\rr)$.
$\forall f \in C(\rr)$, the supp($fg_n$), the closure of 
the domain of $x$ where $f(x)g_n(x)\neq 0$, is in $B_n$ 
since supp$(g_n) \subset B_n$.\ So according to Lemma \ref{lemma1}, 
the elliptic equation
\[
\left\{
\begin{array}{ll}
	(\lambda - {\cal {L}})u &=  fg_{n}\qquad\textrm{in}\quad B_{n}\\
	u|_{\partial B_n}       &=0
\end{array}\right.
\]
has a unique solution $ u_n  \in C^2(B_n) \cap C_0 (B_n)$. 
Thus we can define
\[
	R_n(\lambda)f=\left\{ \begin{array}{ll}
		u_n& \textrm{in} \ B_n	\\
		0 & \textrm{else}.
\end{array}\right.
\]
Furthermore, according to the maximum principle of elliptic operator
(c.f. the proof of Lemma \ref{lemma1}), we have that $R_n(\lambda)$ is 
a positive operator and $\Vert R_n(\lambda)\Vert \leq \frac {1}{\lambda}$.

\begin{lem}
\label{lemma2}
$\forall \lambda > 0$, $f \in C(\rr), \{R_n(\lambda)f\}_{n=1}^{\infty}$ 
converge to a function $\tilde{f}$, satisfying 
$(\lambda- {\cal L})\tilde{f}$ = $f$.  
\end{lem}

\begin{proof}
1) we first prove the lemma when $f \geq 0$.

First, $\forall$ $m, n \in \N$, $m>n$,  $R_m(\lambda)f$ and 
$R_n(\lambda)f$ satisfy the following equations respectively,
\[
\left\{
\begin{array}{ll}
    (\lambda - {\cal L})R_m(\lambda)f=fg_m\quad &\textrm{in}\quad  B_m
\\
		R_m(\lambda)f|_{\partial B_m} = 0,  &
\end{array}\right.
\]
and
\[
\left\{
\begin{array}{ll}
    (\lambda - {\cal L})R_n(\lambda)f=fg_n\quad &\textrm{in}\quad  B_n
\\
		R_n(\lambda)f|_{\partial B_n}=0.    &
\end{array}\right.
\]
Since $g_m > g_n$ and $f$ is positive, $R_m(\lambda)$ satisfies: 
\begin{equation}
\left\{
\begin{array}{ll}
   (\lambda - {\cal L})R_m(\lambda)f=fg_m\geq fg_n\quad &\textrm{in}\quad B_n
\\
		R_m(\lambda)f|_{\partial B_n} \geq 0=fg_n|_{\partial B_n}.&
\end{array}\right.
\label{eq7}
\end{equation}
Using the maximum principle of elliptic equation, the inequality 
in (\ref{eq7}) yields
\[
R_m(\lambda)f|_{B_n} \geq R_n(\lambda)f|_{B_n} .
\]
Thus, $ R_m(\lambda)f \geq R_n(\lambda)f$, i.e. and $R_n(\lambda)f$ increases 
with $n$.  Since $\{R_n(\lambda)\}_{0^{+}}^{^{\infty}}$ have a uniform boundary 
$\frac{1}{\lambda}$, $\lim_{n \rightarrow \infty} R_n(\lambda)f$ exists. 
Let us denote $\tilde{f} =\lim_{n \rightarrow \infty} R_n(\lambda)f$.

	We now need to prove $\tilde{f} \in C^2(\rr)$ and satisfies the 
equation $(\lambda- {\cal {L}})\tilde{f}= f$.  According to the property of 
$R_m(\lambda)f, \forall \varphi \in C_0^{\infty}(B_k)$ (where $k$ is any 
positive integer), the following equation holds:
\[
	\int_{\rr} fg_m\varphi {\rm d}x =\int_{\rr}((\lambda -{\cal {L}})
		(R_m(\lambda)f))\varphi {\rm d}x =
	\int_{\rr} R_m(\lambda)f((\lambda-{\cal {L}^*})\varphi ){\rm d}x
\]
in which $fg_n \uparrow f, R_m(\lambda)f \uparrow \tilde{f}$ as
$n \rightarrow \infty$ and $f, \tilde{f}$ are bounded. According to the 
dominated convergent theorem, we have 
\[
	\int f\varphi {\rm d}x =
	\int \tilde{f}((\lambda-{\cal {L}^*})\varphi ){\rm d}x.
\]
Therefore  $\tilde{f}$ is a weak solution of elliptic equation 
$(\lambda - {\cal L})u=f$. Because $\partial B_k \in C^{\infty}$, 
and $f \in C(\rr)$, the weak solution is also the strong solution.
So $\tilde f|_{ B_k} \in C^2(B_k)$ and 
$(\lambda - {\cal L})\tilde{f}|_{ B_k}=f|_{B_k}$.

	We now let $k \rightarrow \infty$, then $\tilde{f} \in C^2(\rr)$, 
and $(\lambda-{\cal L})\tilde{f}=f$.

2) Using the above result for positive $f$, now consider 
$\forall f \in C(\rr)$.  There exists $f^{+}=\max (f,0),f^{-}=\max (-f,0)$  
such that $ f = f^{+}-f^{-}$; $f^+$, $f^- \in C(\rr)$. 
The linearity of $R_{n}(\lambda)$ leads to 
$R_{n}(\lambda)f=R_{n}(\lambda)f^{+}-R_{n}(\lambda)f^{-}$. 
Since $R_{n}(\lambda)f^{+}$ and $R_{n}(\lambda)f^{-}$ 
have respective limits $R(\lambda)f^{+}$ and $R(\lambda)f^{-}$,  
$R_{n}(\lambda)f$ has limit $R(\lambda)f^{+}-R(\lambda)f^{-}$, and
\[
	(\lambda - {\cal L})(R(\lambda)f^{+} - 
		R(\lambda)f^{-})=f^{+}-f^{-} = f.
\]
Therefore, we can define $R(\lambda)f=R(\lambda)f^{+}-R(\lambda)f^{-}$ 
and complete the proof of Lemma \ref{lemma2}. 
\end{proof}

	Lemma \ref{lemma2} allows us to define 
$R(\lambda)f \triangleq \tilde{f}$.  According to the proof of 
the lemma, the following three properties are all evident.

\noindent
{\bf Proposition 1}. $\forall \lambda >0, R(\lambda): C(\rr) \rightarrow C(\rr)$, 
has the following properties:\\
1. $R(\lambda)$  is a bounded linear operator on $C(\rr)$ and 
$\Vert R(\lambda) \Vert \leq \frac{1}{\lambda}$;\\
2. $R(\lambda)$ has its null space = $\{ 0 \}$;\\
3. $R(\lambda)$\ is positive, i.e. $f \geq 0$ implies $R(\lambda)f \geq 0$.

\noindent
{\bf Remarks:}
According to Dini theorem \cite{KF}, we could conclude 
that $R_n(\lambda)f$ 
uniformly converge to $R(\lambda)f$ on any bounded domain.
This fact will be useful later.

\subsection{\large Banach Space $\hat {C}(\rr)$} 

	Using $R(\lambda)$, we now define $\hat C(\rr)$.  
First, \[H\triangleq span\{R(\lambda_1)\cdots R(\lambda_s)f\
	\mid\ \lambda_1,\cdots,
	\lambda_s>0,s\in \N,f\in C_0(\mathbb{R}^n)\};
\]
then
\[  
\hat C(\rr)\triangleq \overline{H}(closure\  in\  
	norm\  \Vert \cdot \Vert\  of\  C(\rr)).
\]
\begin{lem}
$\forall \lambda >0, R(\lambda)|_{\hat C(\rr)}: \hat C(\rr) \rightarrow \hat C(\rr),\  and\   
C_0(\rr) \subset \hat {C}({\rr}) \subset C(\rr).$
\end{lem}

\begin{proof}
i) If $f \in H$, then according to the definition of $H$, $R(\lambda)f \in H$.
$\forall f \in \hat {C}(\rr)$, there exists $f_{i} \in H$, 
$\Vert f-f_{i}\Vert \rightarrow 0$, as $i \rightarrow +\infty$. Because 
$R(\lambda)$ is a bounded operator, 
$\Vert R(\lambda)f-R(\lambda)f_{i}\Vert \rightarrow 0$ as 
$i \rightarrow +\infty$.  Thus $R(\lambda)f_{i} \in \hat C(\rr)$ leads to  
$R(\lambda)f \in \hat C(\rr)$.

ii) $C_0(\rr) \subset \hat {C}({\rr}) \subset C(\rr)$.

$\forall g \in C_0^{\infty}(\rr)$, and with a compact support, 
$R_{n}(\lambda)(\lambda-{\cal L})g=g$ when $n$ is 
sufficiently large. So  $g=R(\lambda)(\lambda-{\cal L})g \in H$.
Since $\overline {C_0^{\infty}(\rr)}=C_0(\rr)$ and 
$\hat {C}(\rr)=\overline {H}$, we have \ $C_0(\rr) \subset \hat {C}(\rr)$. 
\end{proof}

\subsection{\large Solution to Kolmogorov Equation}

	First, we prove $R(\lambda)$ has the resolvent property, then we 
can define ${\cal D (L)}\triangleq R(\lambda)\hat C(\rr)$ 
which is independent of $\lambda$.

\begin{lem}
\label{lemma4}
$\forall f \in C(\rr)$, $\lambda_1$, $\lambda_2 > 0$,
$R(\lambda_1)f-R(\lambda_2) f$ = 
$(\lambda_2-\lambda_1)R(\lambda_1)R(\lambda_2)f$.
\end{lem}

\begin{proof}
Similar to Lemma \ref{lemma2}, we only need to prove the result when $f \geq 0$.

1) First we show when $\lambda_2 > \lambda_1 > 0$, $\forall n \in N$,
\[
u	(\lambda_2-\lambda_1)R_{n+1}(\lambda_1)R_{n+1}(\lambda_2)f 
	\geq R_{n}(\lambda_1)f-R_{n}(\lambda_2)f 
	\geq (\lambda_2-\lambda_1)R_{n}(\lambda_1)R_{n}(\lambda_2)f.
\]
According to the definition of $R_{n}(\lambda)$,
$(\lambda_2-\lambda_1)R_{n+1}(\lambda_1)R_{n+1}(\lambda_2)$
satisfies the equations
\begin{equation}
\left\{
\begin{array}{ll}
	(\lambda_1- {\cal L})((\lambda_2-\lambda_1)R_{n+1}(\lambda_1)
	R_{n+1}(\lambda_2)f)&=(\lambda_2-\lambda_1)
	(R_{n+1}(\lambda_2)f)g_{n+1}\quad in\quad B_{n+1} \\
	(\lambda_2-\lambda_1)R_{n+1}(\lambda_1)R_{n+1}(\lambda_2)
		f|_{\partial B_{n+1}}&=0;
\end{array}\right.  
\label{eq8}
\end{equation}
$(\lambda_2-\lambda_1)R_{n}(\lambda_1)R_{n}(\lambda_2)$ satisfies the 
equations
\begin{equation}
\left\{
\begin{array}{ll}
	(\lambda_1- {\cal L})((\lambda_2-\lambda_1)R_{n}(\lambda_1)
	R_{n}(\lambda_2)f)&=(\lambda_2-\lambda_1)
	(R_{n}(\lambda_2)f)g_{n}\quad in \quad B_{n} \\
	(\lambda_2-\lambda_1)R_{n}(\lambda_1)R_{n}(\lambda_2)
		f|_{\partial B_{n}}&=0;
\end{array}\right.
\label{eq9}
\end{equation}
$R_{n}(\lambda_1)f-R_{n}(\lambda_2)f$ satisfies the equations
\begin{equation}
\left\{
\begin{array}{ll}
	(\lambda_1- {\cal L})(R_{n}(\lambda_1)f-R_{n}(\lambda_2)f)
	&=(\lambda_2-\lambda_1)R_{n}(\lambda_2)f \quad in \quad B_{n} \\
	(R_{n}(\lambda_1)-R_{n}(\lambda_2))f|_{\partial B_{n}}&=0.
\end{array}\right.
\label{eq10}
\end{equation}
Using the maximum principle of elliptic equation, from (\ref{eq8}), 
(\ref{eq10}), $g_{n+1}|_{B_{n}}$ = 1, and 
$R_{n+1}(\lambda_1)R_{n+1}(\lambda_2)f|_{\partial B_{n}}$ $\geq$ 0, we 
have 
\[
	(\lambda_2-\lambda_1)R_{n+1}(\lambda_1)R_{n+1}(\lambda_2)f 
	\geq R_{n}(\lambda_1)f-R_{n}(\lambda_2)f;
\]
from (\ref{eq9}), (\ref{eq10}), and $0 \leq g_{n} \leq 1$, we have 
\[
	R_{n}(\lambda_1)f-R_{n}(\lambda_2)f \geq 
	(\lambda_2-\lambda_1)R_{n}(\lambda_1)R_{n}(\lambda_2)f.
\]
Thus,
\[
	(\lambda_2-\lambda_1)R_{n+1}(\lambda_1)R_{n+1}(\lambda_2)f 
	\geq R_{n}(\lambda_1)f-R_{n}(\lambda_2)f 
	\geq (\lambda_2-\lambda_1)R_{n}(\lambda_1)R_{n}(\lambda_2)f.
\]

2) Now $\lambda_2 > \lambda_1$, taking limit of the increasing sequence, 
we have $\lim_{n \rightarrow \infty} R_{n}(\lambda_1)R_{n}(\lambda_2)f$ =
$R(\lambda_1)R(\lambda_2)f$.  On one hand,
$\forall \lambda >0$, $n \in N$, $R_{n}(\lambda)f\leq R(\lambda)f$, so 
$R_{n}(\lambda_1)R_{n}(\lambda_2)f$ $\leq$ $R_{n}(\lambda_1)R(\lambda_2)f$
$\leq R(\lambda_1)R(\lambda_2)f$. ($R_{n}(\lambda)$ and $R(\lambda)$
are all positive.) Thus 
$\ \overline{\lim_{n \rightarrow \infty}} R_{n}(\lambda_1)R_{n}(\lambda_2)f$
$\leq$ $R(\lambda_1)R(\lambda_2)f$. 
On the other hand, $\forall\ \ell\in N$, when $n > \ell$, 
$R_{\ell}(\lambda_1)R_{n}(\lambda_2)f$ satisfies the equation
\begin{equation}
\left\{
\begin{array}{ll}
(\lambda_1- {\cal L})(R_{\ell}(\lambda_1)R_{n}(\lambda_2)f)&=
(R_{n}(\lambda_2)f)g_{\ell} \quad  in\quad  B_{\ell} \\
(R_{\ell}(\lambda_1)R_{n}(\lambda_2))f|_{\partial B_{\ell}}&=0;
\end{array}\right.
\label{eq11}
\end{equation} 
$R_{\ell}(\lambda_1)R(\lambda_2)f$\ \  satisfies the equations
\begin{equation}
\left\{
\begin{array}{ll}
(\lambda_1- {\cal L})(R_{\ell}(\lambda_1)R(\lambda_2)f)&=
(R(\lambda_2)f)g_{\ell} \quad  in\quad B_{\ell} \\
(R_{\ell}(\lambda_1)R(\lambda_2))f|_{\partial B_{\ell}}&=0.
\end{array}\right.
\label{eq12}
\end{equation}
According to the remark after the proof of Lemma \ref{lemma2}, in $B_{\ell}$, 
$(R_{n}(\lambda_2)f)g_{\ell}$ uniformly converges to $(R(\lambda_2)f)g_{\ell}$.
Again using maximum principle of elliptic equation, from Eqs. (\ref{eq11})
and (\ref{eq12}), we have 

\[ 
 	|R_{\ell}(\lambda_1)R_{n}(\lambda_2)f(x) 
	- R_{\ell}(\lambda_1)R(\lambda_2)f(x)| \rightarrow 0, \quad 
	n \rightarrow \infty.
\]
Since $R_{\ell}(\lambda_1)R_{n}(\lambda_2)f$ $\leq$ 
$R_{n}(\lambda_1)R_{n}(\lambda_2)f$ where $n > \ell$, there exists
\[ 
     R_{\ell}(\lambda_1)R(\lambda_2)f \leq 
     \lim_{\overline{n\rightarrow \infty}} R_{n}(\lambda_1)R_{n}(\lambda_2)f.
\]
Let $\ell \rightarrow\infty$, 
\[
    R(\lambda_1)R(\lambda_2)f \leq 
    \lim_{\overline{n\rightarrow \infty}} R_{n}(\lambda_1)R_{n}(\lambda_2)f.
\]
Therefore, for $\lambda_2 > \lambda_1$, 
\[
    \lim_{n\rightarrow\infty} R_{n}(\lambda_1)R_{n}(\lambda_2)f
	= R(\lambda_1)R(\lambda_2)f.
\]
From 1) and 2),
\[
	(\lambda_2- \lambda_1)R(\lambda_1)R(\lambda_2)f =
	\lim_{n\rightarrow\infty} (\lambda_2-\lambda_1)R_{n}
	(\lambda_1)R_{n}(\lambda_2)f = R(\lambda_1)f-R(\lambda_2)f.
\]
When $\lambda_1 > \lambda_2$ the same is true.
\end{proof}

From Lemma \ref{lemma4}, $\forall f \in \hat {C}(\rr)$, $\lambda_1,\ \lambda_2\ >0$,
\begin{eqnarray*}
  R(\lambda_2)f &=& R(\lambda_1)f-(\lambda_2-\lambda_1)R(\lambda_1)R(\lambda_2)f \\
                &=& R(\lambda_1)(f-(\lambda_2-\lambda_1)R(\lambda_2)f)
 \in R(\lambda_1)\hat {C}(\rr).
\end{eqnarray*}
So $R(\lambda) \hat {C}(\rr)$ is independent of $\lambda$.  Now we can 
define the domain of ${\cal L}$: ${\cal D}({\cal L})\triangleq
R(\lambda)\hat {C}(\rr)$.

	We now show that the operator ${\cal L}$ on $\hat {C}(\rr)$ 
satisfies the conditions of Hille-Yosida theorem, and complete the
proof of Theorem 1.

\begin{proof}

(1)  ${\cal D}({\cal L})$ is dense in $\hat {C}(\rr)$. Obviously 
$H \subset {\cal D}({\cal L})$, and $H$ is dense in $\hat {C}(\rr)$. So  
${\cal D}({\cal L})$ is dense in $\hat {C}(\rr)$.

(2) $\forall \lambda >0$, $R(\lambda)(\lambda -{\cal L})=E|_{{\cal D}({\cal L})}$,
$(\lambda -  {\cal L})R(\lambda) =E|_{\hat {C}(\rr)}$, where $E$ is the
identity operator.  To prove this, taking any $f \in {\cal D}({\cal L})$, 
according to  ${\cal D}({\cal L})=R(\lambda)\hat {C}(\rr)$, there exists 
$g \in \hat{C}(\rr),\ \ f=R(\lambda)g$.  Thus $R(\lambda)(\lambda -{\cal L})f$ 
= $R(\lambda)(\lambda-{\cal L})R(\lambda)g=R(\lambda)g=f$. 
Then $R(\lambda)(\lambda -  {\cal L})=E|_{{\cal D}({\cal L})}$.
$ (\lambda -{\cal L})R(\lambda)=E|_{\hat {C}(\rr)}$ is a conclusion of Lemma 
\ref{lemma2}.

(3) $\Vert R(\lambda) \Vert \leq \frac{1}{\lambda}$, this is the result of 
Proposition 1.  As the inverse of the bounded operator $R(\lambda)$, 
${\cal L}$\  is close.  

	So the conditions of Hille-Yosida Theorem are all satisfied.
\end{proof}

\noindent
{\bf Remarks:} It is easy to verify from the construction process that 
the solution obtained here is the minimal one.  Uniqueness actually 
does not hold true for general $a_{ij}(x)$ and $b_{i}(x)$.  
Denoting the semigroup obtained by $T(t)$, the solution of 
Kolmogorov-backward equation is $T(t)f$. Next we 
continue to discuss Kolmogorov forward equation and the relation between the 
two solutions.

\begin{Th}
If the coefficients of Eq. (\ref{FPE}) satisfy the assumptions 1), 2) and 3) 
in Introduction, then there exists a Banach space $\widetilde{C}(\rr)$, satisfying 
$C_0(\rr) \subset \widetilde{C}(\rr) \subset C(\rr)$, and 
$\forall g \in \widetilde{C}(\rr)$, the solution of the Cauchy problem (2) and (3)
with initial data g(x) exists in $\widetilde{C}(\rr)$, which is denoted by 
$\tilde{T}(t)g$. Furthermore,\ $\forall f,g \in C_0(\rr)$
\begin{equation}
	\int (T(t)f)g{\rm d}x=\int f(\tilde{T}(t)g){\rm d}x.
\end{equation} 
\end{Th} 

\begin{proof}
One notices that ${\cal L}^{*}u$ contains a term  $\nabla\cdot (b(x)u(x))$, 
so the assumption 2) in Introduction is required to ensure that we can take
${\cal L}^*$ as well as ${\cal L}$ into our consideration. 
For ${\cal L^*}$ we can repeat the steps in the proof of Theorem 1: 
defining the corresponding 
$\tilde{R}_{n}(\lambda)$, $\tilde{R}(\lambda)$ and $\tilde{T}(t)$. 
We only need to prove (13). $R_{n}(\lambda)f$ and $\tilde{R}_{n}(\lambda)g$  
satisfy the following equations respectively.
\[
\left \{
\begin{array}{ll}
(\lambda - {\cal L})R_{n}(\lambda)f&=fg_{n}\ \ \ \ \ \ \ \ in\ \ \  B_{n}\\
R_{n}(\lambda)f|_{\partial B_{n}}&=0
\end{array}\right.
\]
\[
\left \{
\begin{array}{ll}
(\lambda - {\cal L}^{*})\tilde{R}_{n}(\lambda)g&=gg_{n}\ \ \ \ \ \ \ \ in\ \ \   B_{n}\\
\tilde{R}_{n}(\lambda)g|_{\partial B_{n}}&=0.
\end{array}\right.
\]
Then 
\[
\begin{array}{ll}
\int fg_{n}\tilde{R}_{n}(\lambda)g{\rm d}x&=\int_{B_{n}} fg_{n}\tilde{R}_{n}(\lambda)g{\rm d}x \\
&=\int_{B_{n}}((\lambda-{\cal L})R_{n}(\lambda)f)\tilde{R}_{n}(\lambda)g{\rm d}x\\
&=\int_{B_{n}}(R_{n}(\lambda)f)gg_{n}{\rm d}x\\
&=\int (R_{n}(\lambda)f)gg_{n}{\rm d}x
\end{array}
\]
Let $n \rightarrow \infty,\ \ \ \  \int f(\tilde{R}(\lambda)g){\rm d}x=\int (R(\lambda)f)g{\rm d}x$.
According to the theory of Laplace transformation and  the continuity of T(t) and $\tilde{T} (t)$\ , we have
\[ 	
	\int f(\tilde{T}(t)g){\rm d}x=\int (T(t)f)g{\rm d}x.
\]
\end{proof}

\section{\Large Construction of Stationary Markov Process}
\noindent

	In this part, we prove the semigroup constructed in Section 1 has a family 
of transition functions satisfying Kolmogorov-Chapman equation. Then through 
the invariant functional, we find the invariant probability density.  In the end, 
for a general diffusion operator, we obtain its minimal stationary Markov process. 
We first state a simple but important property of the semigroup.

\subsection{\large Transition Functions}

\begin{lem}
	T(t) constructed in Section 1 is a positive semigroup.
\label{lemma5}
\end{lem}

\begin{proof}
We know $R(\lambda)$ is positive (Proposition 1). So the lemma can be 
simply concluded from the relation between $T(t)$ and $R(\lambda)$, $T(t)$ = 
$\lim_{\lambda\rightarrow\infty} e^{-\lambda t}e^{t \lambda ^2 R(\lambda)}f$.
\end{proof}

Having Lemma \ref{lemma5}, when setting $t_0$ and $x_0$\ fixed, we define a 
positive linear functional $\Lambda_{t_0}(x_0)$ on $\hat {C}(\rr)$, 
$\Lambda_{t_0}(x_0)$: $\hat {C}(\rr) \rightarrow {\mathbb R}$, 
$\Lambda_{t_0}(x_0)f$=$T(t_0)f(x_0)$. Restricted on $C_0(\rr)$, 
$\Lambda_{t_{0}}(x_0)|_{C_0(\rr)}$ is also a positive functional. According to 
the Riesz representation theorem, there exists a regular measure 
$p(t_0,x_0,{\rm d}y)$, such that $\forall f \in C_0(\rr)$, 
$\Lambda_{t_0}(x_0)|_{C_0(\rr)}(f)$ = $\int_{\rr}p(t_0,x_0,{\rm d}y)f(y)$.  

Thus, we have the following Lemma.

\begin{lem}
$\forall t>0$, $x \in \rr$, there is  a regular measure $p(t,x,{\rm d}y)$,
satisfying:\\
1) T(t)f(x)=$\int p(t,x,{\rm d}y)f(y)$, $\forall f \in C_0(\rr)$;\\
2) Setting $\Gamma \in {\cal B}$, a Borel field generated by $\rr$, 
$p(t,x,\Gamma)$ is a Borel measurable function.
\label{lemma6}
\end{lem}

\begin{proof}
We have proved the existence of $p(t,x,{\rm d}y)$ and 1). We refer to 
\cite{D}, p.159, for the proof of 2).
\end{proof}

	Because every $f$ in $\hat {C}(\rr)$ is bounded, any $\varphi$  
in ${\cal L}^{1}(\rr)$ can be immersed into  $(\hat {C}(\rr))^{*}$ by
\[
   i:{\cal L}(\rr) \rightarrow (\hat {C}(\rr))^{*};\quad \forall 
	\varphi \in {\cal L}(\rr),\quad f \in \hat {C}(\rr), \quad
	  i(\varphi)(f) \triangleq \int f \varphi {\rm d}x .
\]
Noticing this and identifying $i(\varphi)$\ with $\varphi$\ in the following, we have
\begin{Th}
The transition functions satisfy the Kolmogorov-Chapman equation
\[
	p(t+s,x,\Gamma)=\int p(t,x,{\rm d}z)p(s,z,\Gamma) \quad a.e.
\] 
\end{Th}

\begin{proof}
In $C_0(\rr)$, we have 
\[
	T(t)f(x)=\int p(t,x,{\rm d}y)f(y).
\]
So $\forall f \in C_0(\rr),\ \  \varphi \in {\cal L}^1 (\rr)$,
\begin{eqnarray*}
	T^{*}(t)(\varphi)(f) &=& \int T(t)f \varphi{ \rm d}x 	\\
 &=& \int\left(\int p(t,x,{\rm d}y)f(y)\right)\varphi (x){\rm d}x		\\
 &=& \int\left(\int p(t,x,{\rm d}y)\varphi (x){\rm d}x\right)f(y) \quad 
		\textrm{(Fubini theorem)}.
\end{eqnarray*}
So
\[
	T^{*}(t)(i(\varphi))=\int p(t,x,{\rm d}y)\varphi (x){\rm d}x,
\]
where $T^{*}(t)$ is the conjugate operator of $T(t)$ in $(\hat {C}(\rr))^{*}$.
Then $\forall f \in \hat {C}(\rr)$,
\begin{eqnarray*}
	(T^{*}(t) \varphi )(f) &=& 
	\int \left( \int p(t,x,{\rm d}y)\varphi (x){\rm d}x \right)
			f(y){\rm d}y   \\
   &=& \int \left( \int p(t,x,{\rm d}y)f(y) \right) \varphi (x){\rm d}x
\end{eqnarray*}
So 
\[
	\int (T(t)f)\varphi {\rm d}x =
		\int \int p(t,x,{\rm d}y)f(y)\varphi (x){\rm d}x,
\]
and
\[
	T(t)f(x)=\int p(t,x,{\rm d}y)f(y)\quad  a.e. 
		\quad  \forall f \in \hat{C}(\rr).
\]
From this result, we can prove Kolmogorov-Chapman equation. Because  
$T(t)$ is a semigroup, $T(t+s)f=T(t)(T(s)f)$ $\forall f \in \hat {C}(\rr)$.
\[
\begin{array}{ll}
\int p(t+s,x,{\rm d}y)f(y)&=T(t)(\int p(s,x,{\rm d}y)f(y))\\
&\stackrel {a.e.}{=}\int p(t,x,{\rm d}z)\int p(s,z,{\rm d}y)f(y)\\
&=\int \left(\int p(t,x,{\rm d}z)p(s,z,{\rm d}y)f(y) \right).
\end{array}
\]
Since $f$ is arbitrary,
\[
	\int p(t+s,x,{\rm d}y ) = \int p(t,x,{\rm d}z)p(s,z,{\rm d}y) \quad a.e.
\]
\end{proof}

\begin{co}
Taking the indicator of $B_{n}$:
\[
\begin{array}{ll}
\int X_{B_{n}}p(t,x,{\rm d}y) &\leq \int g_{n+1}p(t,x,{\rm d}y) \\
&=T(t)(g_{n+1})\\
&\leq\ 1.
\end{array}
\]
Let $n\rightarrow \infty$, we have
\[
\int p(t,x,{\rm d}y)\leq 1.
\]
\end{co}

\begin{co}
For $\tilde{T}(t)$, there also exists a family of measure 
$\tilde{p}(t,x,{\rm d}y)$, satisfying the same property as 
$p(t,x,{\rm d}y)$; and 
\begin{equation}
	\tilde{p}(t,x,{\rm d}y){\rm d}x=p(t,y,{\rm d}x){\rm d}y.
\label{eq14}
\end{equation}
\end{co}

\begin{proof}
We only prove (\ref{eq14}). Theorem 2 states that 
$\forall f,g \in C_0^{\infty}(\rr)$
\[
	\int \left( T(t)f \right) g{\rm d}x =
		\int \left( \tilde{T}(t)g \right) f{\rm d}x.
\] 
Thus,
\[
	\int \int p(t,x,{\rm d}y)f(y)g(x){\rm d}x =
		\int \int \tilde{p}(t,y,{\rm d}x)g(x)f(y){\rm d}y.
\]
Since $f$ and $g$ are arbitrary, we have 
\[
	p(t,x,{\rm d}y){\rm d}x=\tilde{p}(t,y,{\rm d}x){\rm d}y.
\]
\end{proof}

\subsection{\large Invariant Functional}

	In the following, we aim at the existence of an invariant 
probability density. The tast is made easier by an indirect approach:
We first construct an invariant functional, and using its Riesz 
representation we arrive at the final goal.
\begin{lem}
$\hat {C}(\rr)$ is a separable space. 
\label{lemma7}
\end{lem}

\begin{proof}
According to the definition of $\hat {C}(\rr)$,
\[
	\hat {C}(\rr) = \overline{span \{R(\lambda_1)\cdots R(\lambda_s)f\ 
	\mid\ \lambda_1, \cdots, \lambda_s>0, s\in N,f\in C_0(\mathbb{R}^n)\}}
\]
we prove that 
\[
	\hat {C}(\rr) = \overline{span\{R(\lambda_1)\cdots R(\lambda_s)f\ 
	\mid\ \lambda_1, \cdots, \lambda_s \in {\cal Q^{+}}, 
	s\in N,f\in C_0(\mathbb{R}^n)\}}
\]
where ${\cal Q}$ is the set of rational number. Then the separability of 
$\hat {C}(\rr)$ becomes obvious from the separability of $C_0(\rr)$.

	Noting the Lemma \ref{lemma4} in Section 1, 
$R(\lambda_1)f-R(\lambda_2)f$ = $(\lambda_2-\lambda_1)R(\lambda_1)
R(\lambda_2)f$,\ $\forall f \in \hat {C}(\rr)$,\  we have 
\[
	\Vert R(\lambda_1)f-R(\lambda_2)f \Vert \leq 
		\vert \lambda_2 - \lambda_1 \vert \Vert 
		R(\lambda_1)\Vert \Vert R(\lambda_2)\Vert \Vert f \Vert.
\]
As $R(\lambda)$\ is a bounded operator and $\cal{Q}^{+}$\ is dense in 
$\mathbb {R^{+}}$,
\[
	\overline{span\{R(\lambda_1)\cdots R(\lambda_s)f\ \mid\ \lambda_1,
	\cdots, \lambda_s \in {\cal{Q}}^{+}, s \in N,f\in C_0(\rr)\}}
	= \hat {C}(\rr). 
\]
So $\hat {C}(\rr)$ is separable.
\end{proof}

\begin{Th}
If $\frac{1}{T} \int_0^{T} T(t)f(x){\rm d}t$\ does not converge to 0 for every 
$f \in C_0(\rr)$, $x \in \rr$, then there exists a positive linear functional 
$\Lambda$ on $\hat {C}(\rr)$, which is invariant under $T(t)$:
$\Lambda (T(t)f)$ = $\Lambda (f)$.  And corresponding to $\Lambda$, there is a
regular measure $\theta ({\rm d}x)$, satisfying
\[
	\int T(t)f(x)\theta ({\rm d}x) \leq \int f(x)\theta ({\rm d}x)
       		\quad f \in C_0(\rr),\ f \geq 0.
\]
Furthermore $\theta ({\rm d}x)$\  has a density ${\theta}(x) >0$.
\end{Th}

\begin{proof}
The proof consists of four parts: 

1) According to our assumption, there exists $f_0 \in C_0(\rr)$, and 
$x_0 \in \rr$
\[
	\overline{lim}_{t\rightarrow \infty} \frac{1}{t} \int_0^{t} 
		T(s)f_0(x_0){\rm d}s \neq 0.
\]
We could assume that $f_0 \geq 0$ (or we can substitute $f_0^{+}$ ($f_0^{-}$)
for $f_0$).  Then there exits $t_{n}$, $t_{n} \rightarrow \infty$, as
$n \rightarrow \infty$, such that
\[
	\lim_{t_n\rightarrow \infty} \frac{1}{t_n} \int_0^{t_n}
	T(s)f_0(x_0){\rm d}s=a > 0.
\] 
The separability of $\hat {C}(\rr)$ provides us a sequence of functions 
$\{f_{n} \}_{n=1}^{^{\infty}}$, which is dense in $\hat {C}(\rr)$. 
Because $\Vert T(s)f(x) \Vert \leq \Vert f \Vert$,\ $T(s)f_{k}(x)$ have a 
bound independent of $s$.  Using the critical Cantor-diagonal method, we 
could choose a subsequence of ${t_n}$, which is still denoted by 
${t_{n_{i} }}$, such that $\forall f_{k}$
\[
	\lim_{i\rightarrow\infty} \frac{1}{t_{n_{i}}} \int_0^{t_{n_{i}}}
	T(s)f_{k}(x_0){\rm d}s  \qquad \textrm{exists}.
\]
Having $\{f_{k}\}_{k=1}^{^{\infty}}$ \ dense in $\hat{C}(\rr)$, we get 
$\forall f \in \hat {C}(\rr)$
\[
	\lim_{k\rightarrow\infty} \frac{1}{t_{n_{k}}} \int_0^{t_{n_{k}}}
	T(s)f(x_0){\rm d}s \quad \textrm{exists}.
\]
Now define $\Lambda$: $\hat {C}(\rr) \rightarrow \mathbb {R}$
\[
	\Lambda (f) = \lim_{k\rightarrow\infty} \frac{1}{t_{n_{k}}} 
		\int_0^{t_{n_{k}}}T(s)f(x_0){\rm d}s.
\]
The assumption makes sure that $\Gamma$ is not zero and it is straightforward 
to prove that $\Lambda \in (\hat {C}(\rr))^{*}$, and 
$\vert \Lambda (f)\vert \leq \Vert f\Vert$\ $\forall f \in \hat C(\rr)$.

2) $\Lambda (T(t)f)=\Lambda(f)$\ $\forall f \in \hat {C}(\rr)$.
\begin{eqnarray*}
    \Lambda (T(t)f) &=& \lim_{k\rightarrow\infty} \frac{1}{t_{n_{k}}} 
			\int_0^{t_{n_{k}}}T(s)(T(t)f)(x_0){\rm d}s	\\
             &=& \lim_{k\rightarrow\infty} \frac{1}{t_{n_{k}}} 
			\int_0^{t_{n_{k}}}T(s+t)f(x_0){\rm d}s		\\
	     &=& \lim_{k\rightarrow\infty} \frac{1}{t_{n_{k}}} 
			\int_{t}^{t+t_{n_{k}}}T(s)f(x_0){\rm d}s	\\
	     &=& \lim_{k\rightarrow\infty} \left\{\frac{1}{t_{n_{k}}} 
			\left(\int_0^{t_{n_{k}}}T(s)f(x_0){\rm d}s +
			\int_{t_{n_{k}}}^{t_{n_{k}}+t}T(s)f(x_0){\rm d}s i
		      - \int_0^{t}T(s)f(x_0){\rm d}s \right)\right\}.
\end{eqnarray*}
Since $\left|\int_{t_{n_{k}}}^{t_{n_{k}}+t}T(s)f(x_0){\rm d}s\right|$ 
$\leq t \Vert f \Vert$, $\left|\int_0^{t}T(s)f(x_0){\rm d}s\right|$ 
$\leq t\Vert f\Vert$,
\[
     \Lambda (T(t)f)=\lim_{k\rightarrow\infty}
	\frac{1}{t_{n_{k}}}\int_0^{t_{n_{k}}}T(s)f(x_0){\rm d}s=\Lambda (f).
\]

3) Restricted on $C_0(\rr)$,\ $\Lambda|_{C_0(\rr)}$\ is still a positive linear 
functional on $C_0(\rr)$.\ According to Riesz representation theorem, there exists 
a regular measure $\theta ({\rm d}x)$,\  such that 
\[
    \Lambda (f)=\int f(x)\theta ({\rm d}x),\quad \forall f \in C_0(\rr).
\]
Now  $\forall f \in C_0(\rr),\ f \geq 0$,\ we have\ $(T(t)f)g_{n} \in C_0(\rr)$.\
($g_{n}$\ is defined in the Section 1.2),\ and 
\[
	\Lambda((T(t)f)g_{n})=\int (T(t)f)g_{n} \theta ({\rm d}x).
\]
As $g_{n} \uparrow \ 1,\ \ $ and $\Lambda $ is positive, we have
\[
	\Lambda(T(t)f) \geq \lim_{n\rightarrow\infty} \Lambda (T(t)fg_{n})
		=\int T(t)f\theta ({\rm d}x).
\]
From 2),\ $ \Lambda (T(t)f)=\Lambda (f)=\int f\theta ({\rm d}x)$, we get  
$\int T(t)f \theta ({\rm d}x) \leq \int f \theta({\rm d}x),\ 
where f \in C_0(\rr).$

4) $\theta ({\rm d}x)$\  has a density $\theta (x)>0$.\\
As $\Lambda$ is an invariant functional, using $R(\lambda)$ =
$\int_0^{\infty} e^{-\lambda t} T(t){\rm d}t$ \ and the method in 
\cite{LM}, we have 
\[
     \lambda \Lambda[R(\lambda)f]=\Lambda (f),\quad \forall f \in \hat {C}(\rr).
\]
And from $R(\lambda)(\lambda-{\cal L})f=f$, $\forall f \in C_0^{\infty}(\rr)$,\ 
take $\lambda =1$,\ we have $\Lambda ({\cal L}f)$=0,\ and\  
$\int{\cal L}f \theta ({\rm d}x)=0$. So\  $\theta({\rm d}x)$\ is the weak 
solution of ${\cal L}^{*} \theta=0$ \  in the space ${\cal D}_0(\rr)$\ of 
generalized functions.\  According to Schwartz-Weyl lemma, there exists an
infinitely differentiable function $\theta (x) \in {\cal L}^1(\rr)$,\ such that 
$\theta (\Gamma)=\int_{\Gamma} \theta(x){\rm d}x$.

Because $\Lambda \neq 0,\ \theta (x) \geq 0$,\ using the strong maximum principle of elliptic equation,
we have $ \theta(x)>0$.
\end{proof}

\begin{co}$\forall f \in C_0(\rr)$
\begin{eqnarray*}
	\int T(t)f \theta ({\rm d}x) &=&
		\int\left(\int p(t,x,{\rm d}y)f(y)\right)\theta ({\rm d}x)   \\
	&=& \int\left(\int p(t,x,{\rm d}y)\theta ({\rm d}x)\right)f(y)
		\quad \textrm{(Fubini Theorem)}		  	  \\
	&\leq& \int f(y)\theta ({\rm d}y).
\end{eqnarray*}
Since $\theta ({\rm d }y)$\ is regular and $f$ is arbitrary,  we have 
\[
	\int p(t,x,{\rm d}y)\theta({\rm d}x) \leq \theta ({\rm d}y).
\]
\end{co}

\subsection{\large Invariant Probability}

	In the following, we prove $\theta$ is just the invariant density
\[
	\int p(t,x,{\rm d}y)\theta ({\rm d}x)=\theta ({\rm d}y).
\]
First, we define $e(t,x)\triangleq\int p(t,x,{\rm d}y)$. 
As $p(t,x,{\rm d}y)$\ is a finite measure on $\rr$, $e(t,x)$ is well 
defined and we have the following property of $e(t,x)$.

\begin{lem}
\label{lemma8}
e(t,x) decreases as t $\rightarrow \infty$.
\end{lem}

\begin{proof} 
According to the Kolmogorov-Chapman equation,
\[
	p(t+s,x,{\rm d}y)=\int p(t,x,{\rm d}z)p(s,z,{\rm d}y).
\]
When $s>0$,
\[
\begin{array}{ll}
    e(t+s,x) &=\int p(t+s,x, {\rm d}y)\\
	     &= \int \int p(t,x,{\rm d}z)p(s,z,{\rm d}y)\\
	     &\leq \int p(t,x,{\rm d}z)\quad \textrm{(Corollary 3)} \\
	     &=e(t,x).
\end{array}
\]
Thus, $e(t+s) \leq e(t,x)$.
\end{proof}

Now we define $e(x)\triangleq \lim_{t\rightarrow\infty}e(t,x)$\
$\forall x \in \rr$.  

\begin{lem}
\label{lemma9}
$\int p(t,x,{\rm d}y)e(y)=e(x)$\ $\forall t>0$,\ $x \in \rr$. And under the 
condition of Theorem 4,\ $e(x)>0$\ and satisfies
${\cal L}^{*}e=0.$
\end{lem}

\begin{proof}
1) $\int p(t,x,{\rm d}y)e(y)=e(x)$.\\
According to definition,
\[
\begin{array}{ll}
	e(t+s,x) &=\int p(t+s,x,{\rm d}y)=\int \int p(t,x,{\rm d}z)p(s.z.{\rm d}y)			\quad \textrm{(Theorem 3)}		\\
		 &=\int p(t,x,{\rm d}z)e(s,z).
\end{array}
\]
Let $s \rightarrow \infty$,\ $e(x)$ = $\int p(t,x,{\rm d}z)e(z)\quad$ 
(Levi theorem \cite{KF}).

2) $e(x) \geq 0 \ \ \ \ \ and\ \ \  e(x)\not\equiv 0.$\\
It's obvious that $e(x) \geq 0$. Now we assume $e(x) \equiv 0$.\
From $T(t)f(x)=\int p(t,x,{\rm d}y)f(y)\ \\ \ \forall f \in C_0(\rr)$,\ \ if $\Vert f(x) \Vert \leq 1$, \ then $T(t)f(x) 
\leq e(t,x)$. So the assumption that $e(x) \equiv 0$ leads to $T(t)f(x) \rightarrow 0,\ \ as\ \  t \rightarrow \infty\ \ \ \forall x \in \rr$,
which contradicts the
condition of this lemma. Therefore e(x)$\not\equiv 0$.

3) ${\cal L}e(x)=0,\ \ e(x)>0$.\\
In the Corollary 2, we have proved $\tilde{p}(t,x,{\rm d}y){\rm d}x$ =
$p(t,y,{\rm d}x){\rm d}y$. So
\begin{eqnarray*}
\int e(x)(\tilde{T}(t)g(x)){\rm d}x &=& 
		\int e(x)\int \tilde{p}(t,x,{\rm d}y)g(y){\rm d}x\\
&=& \int \int e(x)g(y)p(t,x,{\rm d}y)\\
&=& \int g(y)e(y){\rm d}y \quad \textrm{(Fubini theorem)}.
\end{eqnarray*}
According to the relation between $\tilde{T}(t)$ and $\tilde{R}(\lambda)$, we have
\[
\lambda \int e(x)(\tilde{R}(\lambda)g(x)){\rm d}x=\int g(x)e(x){\rm d}x.
\]
So 
\[
\int e(x){\cal L}^{*}g {\rm d}x= 0.
\]

Thus $e(x)$ is a solution of ${\cal L}u=0$\ in the generalized sense. And the same 
reasoning as for $\theta (x)$\ leads to $e(x)$ being infinitely differentiable and 
$e(x) > 0$.
\end{proof}

\begin{Th}
Under the conditions of Theorem 4,\ $\theta (x)$ is invariant under T(t).
\end{Th}

\begin{proof}
From the part 1) of Lemma \ref{lemma9},\ 
$\int p(t,x,{\rm d}y) \theta (x) e(y){\rm d}x$ = $\int \theta (x)e(x){\rm d}x$,\  
$e(x)>0$, and $\int p(t,x,{\rm d}y)\theta (x){\rm d}x \leq \theta ({\rm d}y)$, 
we have 
\[
	\int p(t,x,{\rm d}y)\theta (x){\rm d}x
	= \theta ({\rm d}y)\quad \forall t >0,\quad  x \in \rr
\]
Which means $\theta ({\rm d}x)$ is invariant under $T(t)$.
\end{proof}

Now, we have proved that $T(t)$ has a family of transition functions 
$p(t,x,{\rm d}y)$, and an invariant measure $\theta ({\rm d}y)$. Thus  
we can construct a stationary Markov process by Kolmogorov theorem, 
whose transition probability functions are $\{\tilde{p}(t,x,{\rm d}y)\}$ 
and the initial distribution is $\theta ({\rm d}x)$.

\begin{co}
Theorem 4 and 5 together actually shows a weak form of the Foguel 
alternative given in \cite{LM} where diffusions with bound 
coefficients $A(x)$ and $b(x)$ are considered.
\end{co}

\section{\Large Reversibilty  and  Entropy  Production}
\noindent

	With the defintions for time-reversibility and entropy production 
rate, we establish the following equivalence for the diffusion process 
we constructed.
\begin{Th}
For the stationary process constructed in Section 2,\ the following 
three statements are equivalent:\\
(i) The process is time-reversible;\\
(ii) Its corresponding elliptic operator $ {\cal L}^{*}$ is symmetric 
on $C_0^{\infty}(\rr)$\  with respect to a positive function 
$w^{-1}(x) \ , w(x) \in L^1(\rr)$, i.e., $\int_{\rr} w(x)dx < \infty$;\\ 
(iii) The process has zero entropy production rate (epr).
\end{Th}

\begin{proof}
$(i) \Longrightarrow (ii)$.\\
The proof of this result for a discrete state Markov process
is due to Kolmogorov.  According to the definition of reversibility, 
we have $\forall A,B \in{\cal B }$
\[
\int_B \int_A \tilde{p}(t,x,{\rm d}y)\theta(y){\rm d}x
	= \int_A\int_B\tilde{p}(t,y,{\rm d}x)\theta(x){\rm d}y,
\]
\[
\int_B\int_A\chi_{_{A}}(x)\tilde{p}(t,x,{d}y)\chi_{_{B}}(y)
	\theta(y){\rm d}x =\int_A\int_B \chi_{_{A}}(y)
	\tilde{p}(t,y,{\rm d}x)\chi_{_{B}}(x)\theta(x){\rm d}y.
\]
By the standard method in probability, we have 
\begin{equation}
\int_B\int_A \phi(x) \tilde{p}(t,x,{\rm d}y)\psi(y)
	\theta(y){\rm d}x=\int_A\int_B\psi(y)\tilde{p}(t,y,{\rm d}x)
	\phi(x)\theta(x){\rm d}y
\label{phippsi}
\end{equation}
where $\phi(x)$, $\psi(x) \in C_0^{\infty}(\rr)$. Noting the definition 
of $\tilde{T}(t)$, we differentiate both sides of (\ref{phippsi})  
with respect to $t$ at $t=0$, we have
\[
\int_{\rr}\phi (x){\cal L}^{*}[\theta(x) \psi(x)]{\rm d}x
	=\int_{\rr}\psi(y){\cal L}^{*}[\theta(y) \phi (y)]{\rm d}y .  
\] 
Let $f(x)=\theta(x) \phi (x)$ and $g(x)=\theta (x) \psi(x)$, then $f$ 
and $g$ are two arbitrary functions in $C_0^{\infty}(\rr)$. Since 
$\theta(x) >0$,
\[
   \int_{\rr} \theta^{-1}(x)f(x){\cal L}^{*}[g(x)]{\rm d}x
	=\int_{\rr} \theta^{-1}(y)g(y){\cal L}^{*}[f(y)]{\rm d}y.
\]
Therefore, the operator ${\cal L}^{*}$ is symmetric with respect to 
the reciprocal of its stationary distribution $\theta(x)$: 
$w(x)=\theta(x)$.  This result is known  to physicists.

\noindent
$(ii) \Longrightarrow (iii)$.\\
The differential operator ${\cal L}^{*}$ can also be rewritten as 
\[
	{\cal L}^{*}f=\frac{1}{2}\nabla\cdot(A\nabla f)+(\nabla f)
	\cdot b(x)+ f \nabla\cdot b(x).
\]
The statement ($ii$) is
\[
	\int e^{U}g(x){\cal L}^{*}[f(x)]{\rm d}x
	=\int e^{U}f(x){\cal L}^{*}[g(x)]{\rm d}x,
\] 
in which the positive $w(x)=e^{-U}$, $f$ and $g$ $\in C_{0}^{\infty}(\rr)$ 
are arbitrary functions. This leads to 
\[
	\int e^{U}g\left(\frac{1}{2} \nabla \cdot (A\nabla f)
	+(\nabla f)\cdot b(x)\right){\rm d}x =
	\int e^{U} f\left(\frac{1}{2}\nabla \cdot 
	(A \nabla g)+(\nabla g)\cdot b(x)\right){\rm d}x.
\]
Through integration by part, the first term on the left-hand-side 
(and similarly for the right-hand-side)
\[
    \int e^{U}g\nabla \cdot (A \nabla f){\rm d}x
	=-\int e^{U}(\nabla g)A(\nabla f){\rm d}x
	 -\int e^{U}g(\nabla U)A(\nabla f){\rm d}x, 
\]
and we have
\[
  \int e^{U}g\left(\frac{1}{2}(\nabla U)A(\nabla f)- (\nabla f) 
	\cdot b(x)\right){\rm d}x=\int e^{U}f\left(\frac{1}{2}(\nabla U)
	A(\nabla f)-(\nabla g)\cdot b(x)\right){\rm d}x.
\]
By a simple rearrangement, we have 
\[
	\int e^{U}(g\nabla f-f\nabla g)\cdot
	(\frac{1}{2}A\nabla U-b(x)){\rm d}x=0.
\]
Since $f$ and $g$ are arbitrary, we have $\frac{1}{2} A \nabla U - b(x)=0 $ 
in which $U=-\log w $. Therefore
\[
	\nabla \log w(s)+2A^{-1}b(x)=0,
\]
which means epr=0.

\noindent
$(iii) \Longrightarrow (i)$.\\
The statement epr=0 leads to $\frac{1}{2}A\nabla \theta (x)+b(x)\theta (x)=0$ 
and we know
\begin{equation}
	{\cal L }^{*}\theta = 
	\nabla \cdot (\frac{1}{2}A \nabla \theta +b(x) \theta )=0.
\label{lstareq0}
\end{equation}
Under these conditions, the operators 
$R_{n}(\lambda), R(\lambda), \tilde{R}_{n}(\lambda), \tilde{R}(\lambda)$
have the following properties:

	First, $\theta R_{n}(\lambda)(\psi)$ =
$\tilde{R}_{n}(\lambda)(\theta \psi)$, where 
$\varphi \in C_0^{\infty}(\rr)$.  This is becasue
\[
\begin{array}{ll}
{\cal L}^{*}(\theta R_{n}(\lambda)(\psi))&=\frac{1}{2}\theta\nabla\cdot A\nabla(R_{n}(\lambda)(\psi))-\theta b(x)\cdot \nabla (R_{n}(\lambda)(\psi))\\
&+\frac{1}{2}R_{n}(\lambda)(\psi)\nabla\cdot A\nabla \theta +R_{n}(\lambda)(\psi)b(x)\cdot \nabla \theta +R_{n}(\lambda)(\psi)\theta\nabla \cdot b(x)\\
&+\nabla (R_{n}(\lambda)(\psi))\cdot A\nabla \theta +2\theta \nabla (R_{n}(\lambda)(\psi))\cdot b(x)\\
&=\theta {\cal L}(R_{n}(\lambda)(\psi))+R_{n}(\lambda)(\psi){\cal L}^{*}(\theta)+(\nabla \cdot R_{n}(\psi))(A\nabla \theta+2\theta b(x)) .
\end{array}
\]
Equations (\ref{lstareq0}) leads to 
\[
	{\cal L}^{*}(\theta R_{n}(\lambda)\psi)
		=\theta {\cal L}R_{n}(\lambda)(\psi) .
\]
Thus $\theta R_{n}(\lambda)\psi$\  satisfies 
\[ 
 \left \{
 \begin{array}{ll}
 (\lambda -{\cal L}^{*})(\theta R_{n}(\lambda)\psi)&
	= \theta (\lambda- {\cal L})R_{n}(\lambda)(\psi)
	=\theta \psi g_{n}\quad  in \quad B_{n}\\
 \theta R_{n}(\lambda)(\psi)|_{\partial B_{n}}&=0 .
  \end{array}\right.
\]
According to the uniqueness of the solution in $B_{n}$,
\begin{equation}
	\theta R_{n}(\lambda)(\psi)
		=\tilde{R}_{n}(\lambda)(\theta \psi) .
\label{thetaR}
\end{equation}

	Second, from (\ref{thetaR}), 
$\forall \varphi,\ \psi \in C_0^{\infty}(\rr)$ 
\[
\begin{array}{ll}
\int \psi g_{n}\tilde{R}_{n}(\lambda)(\theta \varphi){\rm d}x&=\int_{B_{n}}\psi g_{n}\theta R_{n}(\lambda)(\varphi){\rm d}x\\
&=\int_{B_{n}}(\lambda-{\cal L}^{*})\tilde{R}_{n}(\lambda)(\psi \theta)R_{n}(\lambda)(\varphi){\rm d}x\\
&=\int_{B_{n}}\tilde{R}_{n}(\lambda)(\psi\theta)(\lambda-{\cal L})(R_{n}(\lambda)(\varphi)){\rm d}x\\
&=\int_{B_{n}}\tilde{R}_{n}(\lambda)(\psi\theta)\varphi g_{n}{\rm d}x\\
&=\int_{B_{n}}\tilde{R}_{n}(\lambda)(\psi \theta)\varphi g_{n}{\rm d}x .
\end{array}
\]
Let $ n \rightarrow \infty $, \ since  $\psi,\ \varphi $ \   are compact supported,
\[
\int \psi \tilde{R}(\lambda)(\theta\varphi){\rm d}x=\int \varphi \tilde{R}(\lambda)(\theta \psi){\rm d}x.
\]
According to the theory of Laplace transformation, from the fact that 
$\psi \tilde{T}(t)(\theta\varphi)$,  and $\varphi \tilde{T}(t)(\theta \psi)$
are continuous with $t$, we have 
\[
\int \psi \tilde{T}(t)(\theta\varphi){\rm d}x=\int \varphi \tilde{T}(t)(\theta \psi){\rm d}x.
\]
This leads to 
\[
\int \int \psi (x)\tilde{p}(t,x,{\rm d}y)\theta (y)\varphi (y){\rm d}x=\int \int \varphi (y) \tilde{p}(t,y,{\rm d}x)\theta (x)psi (x){\rm d}y.
\]
The standard method of measure theory leads to
\[
\int_{A} \int_{B} \tilde{p}(t,x, {\rm d}y)\theta (y){\rm d}x=\int_{B}\int_{A}
     \tilde{p}(t,x,{\rm d}y)\theta (y){\rm d}x
\]
which means reversibility.
\end{proof}

\noindent
{\bf Remarks:}  The symmetric operator in ($ii$) is also 
maximum on an appropriate Hilbert space constructed from 
$\hat{C}(\rr)$;  hence it is self-adjoint. 

We have now come to the conclusion of this work in which we have provided
the general diffusion processes defined by nonlinear stochastic differential
equations (\ref{SDE}) with a sound thermodynamic structure.  We have
introduced two fundamental physical concepts, time-reversibility 
and entropy production, and have shown the equivalence between the 
reversibility and zero entropy production.  We have established mathematically
the essential properties of fluctuating isothermal equilibrium systems.
In a separated report, we shall resume the investigation on the 
asymptotic property of the diffusion processes we constructed and a 
strong form of Foguel alternatives for the general diffusion equation.  
Similar results for the restrictive case of linearly increasing 
$a_{ij}(x)$, and $b_{i}(x)$ have been given in \cite{LM}, which takes
the advantage of the existence of the fundamental solution. Finally, 
in a recent work on certain non-Markovian Gaussian processes \cite{Q4}, it 
has been suggested that the equivalence between time-reversibility and 
equilibrium requires some additional conditions.  A rigorous mathematical 
treatment of this problem remains to be developed.


\begin{thebibliography}{000}

\bibitem{DE} Doi, M., Edwards, S.F.: {\sl The Theory of Polymer
Dynamics.} Oxford: Clarendon Press, 1986. 

\bibitem{D} Dynkin, E.B.: {\sl Markov Process}, New York: 
Springer-Verlag, 1965

\bibitem{ECM} Evans, D.J., Cohen, E.G.D., Morriss, G.P.:
Probability of second law violations in shearing steady-states.
Phys. Rev. Lett. {\bf 71}, 2401-2404. 

\bibitem{FK} Fisher, M.E., Kolomeisky, A.B.: The force exerted
by a molecular motor. Proc. Natl. Acad. Sci. USA {\bf 96}, 
6597-6602 (1999)

\bibitem{GQW} Guo, M.-Z., Qian, M., Wang, Z.D.: The entropy 
production and circulation of diffusion processes on manifold. 
Chin. Sci. Bull. {\bf 42}, 982-985 (1998)

\bibitem{JQQ} Jiang, D.-Q., Qian, M., Qian, M.-P.: Entropy 
production and information gain in axiom-A systems. 
Comm. Math. Phys. {\bf 214}, 389-409 (2000)

\bibitem{Ka} Kalpazidou, S.: {\sl Cycle Representation of Markov Process},
New York: Springer-Verlag, 1994

\bibitem{KF} Kolmogorov, A.N., Fomin, S.V.: {\sl Introductory 
Real Analysis}, New York: Dover, 1970

\bibitem{LM} Lasota, A., Mackey, M.C.: {\sl Chaos, Fractals, and Noise:
Stochastic Aspects of Dynamics}, New York: Springer-Verlag, 1994

\bibitem{LS} Lebowitz, J.L., Spohn, H.: A Gallavotti-Cohen-type
symmetry in the large deviation functional for stochastic dynamics.
J. Stat. Phys. {\bf 95}, 333-365 (1999) 

\bibitem{N} Nelson, E.: An existence theorem for second order 
parabolic equations. Tran. Amer. Math. Soc. {\bf 88}, 414-429 (1958)

\bibitem{NP} Nicolis, G., Prigogine, I.: {\sl Self-organization
in nonequilibrium systems}, New York: Wiley-Interscience, 1977
 
\bibitem{O} Onsager, L.: Reciprocal relations in irreversible 
processes I. Phys. Rev. {\bf 37}, 405-426 (1931)

\bibitem{Q1} Qian, H.: Vector field formalism and analysis for a 
class of Brownian ratchets. Phys. Rev. Lett., {\bf 81}, 3063-3066 (1998)

\bibitem{Q4} Qian, H.: Single-particle tracking: Brownian dynamics of 
viscoelastic materials. Biophys. J., {\bf 79}, 137-143 (2000)

\bibitem{Q2} Qian, H.: Equations for stochastic macromolecular mechanics 
of single proteins: equilibrium fluctuations, transient kinetics and 
nonequilibrium steady-state. physics/0007017. 

\bibitem{Q3} Qian, H.: Mathematical formalism for isothermal linear 
irreversibility. Proc. R. Soc. A. {\bf 457}, 1645-1655 (2001) 

\bibitem{Q5} Qian, H.: Nonequilibrium steady-state circulation and 
heat dissipation functional. Phys. Rev. E. {\bf 64}, 022101 (2001).  

\bibitem{Q6} Qian, H.: Mesoscopic nonequilibrium thermodynamics of 
single macromolecules and dynamic entropy-energy compensation.
Phys. Rev. E., {\bf 65}, 016102 (2002). 

\bibitem{QM1} Qian, M.: Extension of an elliptic differential 
operator and $\widehat{\widehat C}$ semigroups (Chinese).
Acta Math. Sin., {\bf 22}, 471-486 (1979).

\bibitem{QM2} Qian, M.: The invariant measure and ergodic 
property of a Markov semigroup (Chinese). Beijing Daxue Xuebao,
{\bf 2}, 46-59 (1979).

\bibitem{QW} Qian, M., Wang, Z.-D.: The reversibility, entropy 
production and rotation numbers of diffusion processes on compact 
Riemannian manifolds. Comm. Math. Phys. {\bf  206}, 429-445 (1999)

\bibitem{QQ1} Qian, M.-P., Qian, M.:  Circulation for recurrent 
Markov chain. Z. Wahrsch. Verw. Gebiete. {\bf 59}, 203-210 (1982)

\bibitem{QQ2}  Qian M.-P., Qian, M.: The entropy production and 
irreversibility of Markov processes. Chin. Sci. Bull. {\bf 30},
445-447 (1985)

\bibitem{QQG} Qian, M.-P., Qian, M., Gong, G.-L.:  The reversibility and 
the entropy production of Markov processes. Contemp. Math.  {\bf 118},
255-261 (1991)

\bibitem{R} Ruelle, D.: Positivity of entropy production in
the presence of a random thermostat. J. Stat. Phys. {\bf 86},
935-951 (1997)

\bibitem{SV} Stroock, D.W., Varadhan, S.R.S.: {\sl Multidimensional
diffusion processes}, Springer-Verlag, 1979.

\bibitem{Y} Yosida, K.: {\sl Functional Analysis}, 5th Ed., 
Berlin: Springer-Verlag, 1978

\end{thebibliography}
\end{document}